\documentclass[prl, amsfonts, twocolumn, nofootinbib, showpacs]{revtex4}
\usepackage{graphicx, epsfig}

\newcommand{\be}{\begin{equation}}
\newcommand{\ee}{\end{equation}}
\newcommand{\bea}{\begin{eqnarray}}
\newcommand{\eea}{\end{eqnarray}}

\def\etal{{\it et al.}}

\begin{document}

\title{Universal Limits on Computation}

\author{Lawrence M. Krauss$^{1}$} 
\author{Glenn D. Starkman$^{1,2}$}
\affiliation{$^1$ Center for Education and Research in Cosmology and Astrophysics,
Department of Physics, and Department of Astronomy, Case Western Reserve University, Cleveland, OH~~44106-7079 \\
 $^2$ Department of Physics, CERN, Theory Division,
             1211 Geneva 23, Switzerland}

\begin{abstract}

The physical limits to computation have been under active scrutiny over the past decade or two, 
as theoretical investigations of the possible impact of quantum mechanical processes on computing have begun to 
make contact with realizable experimental configurations.   
We demonstrate here that the observed acceleration of the Universe can produce a universal limit on 
the total amount of information that can be stored and processed in the future, 
putting an ultimate limit on future technology for any civilization, including a time-limit on Moore's Law.  
The limits we derive are stringent, and include the possibilities that 
the computing performed is either distributed or local.  
A careful consideration of the effect of horizons on information processing is necessary for this analysis, 
which suggests that the total amount of information that can be processed by any observer 
is significantly less than the Hawking-Bekenstein entropy 
associated with the existence of an event horizon in an accelerating universe. 
\end{abstract}

\maketitle

The physical nature of computation\cite{Landauer} implies fundamental limits 
on the amount of information processing that finite physical systems can perform\cite{Lloyd2000}, 
and even the amount of information that can have been processed since the Big Bang\cite{Lloyd2002}.  
At the same time, it has been recognized recently that the apparent acceleration of the Hubble expansion 
can put fundamental limits on the observable Universe, even infinitely far into the future\cite{Krauss}.  
The observed expansion seems to be consistent with that expected if the energy density is dominated by vacuum energy, 
corresponding to the existence of a cosmological constant\cite{Reiss}.  
In this case, if one treats consciousness, conservatively, as merely a form of computation, 
then one can derive a finite total lifetime for any civilization in an accelerating universe\cite{Krauss}  (see also (\cite{vilenkin,freese}) for ancillary discussions).  
This conclusion results from the fact that 
in such a universe one ultimately has access to only a finite volume, even after an infinite time.  
In the case of actual conscious living systems, it is difficult to quantify the nature of this limit, 
because we do not currently understand the precise relationship between computational complexity and consciousness.  
In a broader sense, however, one can use the two factors stressed above --
the physical nature of computation and the fundamental limits on the accessible volume in an accelerating universe --
to demonstrate that a new universal limit exists on computation itself within the Universe we apparently inhabit.
A universe dominated by vacuum energy asymptotically approaches a de Sitter phase, 
where the scale factor of the universe $a(t)$ 
(associated with the distance between distant galaxies) increases exponentially
\be
\label{deSitter}
a(t) = a_0 e^{H t} .	
\ee
Without loss of generality one can choose $t=0$ at the present time, 
when the scale factor has some fixed magnitude $a_0$.  
The Hubble expansion rate, $H$, is constant in the case of vacuum energy domination (i.e. a cosmological constant). 
In such a de Sitter space there exists a global horizon, representing the maximum distance that any observer can probe.   
Since all distant objects recede from the observer at a rate proportional to their distance, 
objects greater than a specific distance, called the horizon distance and given by $c/H$, 
will be receding at a velocity that exceeds the speed of light, $c$.  
They will thus cease to be in causal contact with the observer.
One may use this fact to determine two fundamental parameters:
the total energy contained within the horizon, and the total entropy.  
One might imagine that these will provide direct upper limits 
on the total amount of information processing possible in the future of the Universe.  
However, one must consider the actual limits somewhat more carefully.  
In the first place, not all of this energy is accessible to an observer located at some point within this volume.  
Due to the Hubble expansion, the later the observer attempts to collect energy within the accessible volume,
the less of it there is.  
Moreover if one sends out machines to collect energy 
and if they then beam back this energy to the observer in the form of radiation,
then the energy density will redshift during its journey.  
Alternatively, if one considers a network of distributed computers 
processing information in different locations before beaming results back to a central processor, 
these computers will only have a finite time to perform their operations before they lose causal contact with this processor.  

We first consider the total amount of energy that one can harvest centrally.  
Assume for simplicity that the expansion is pure de Sitter, as given in equation \ref{deSitter}.  
In this case one can define a conformal time parameter  
\be
\label{conformaltime}
\eta(t) = \int_0^t\frac{dt'}{a(t')} , 
\ee
in terms of which most calculations can be most easily performed.  
In terms of this parameter:
\be
a(\eta) = \frac{a_0}{1 - a_0 H \eta}
\ee

To determine the maximum accessible total energy that can be collected and stored, 
imagine sending out a spherical shell of self-replicating devices 
each of which can deposit an instruction set for locally building machines 
to convert rest mass into radiation energy and beam it back to the central observer.  
(Since the total radiation energy density today is so small, 
harvesting radiation is negligible compared to harvesting matter.) 
For example, these machines might produce mini-black holes from background material.  
These black holes would then quickly evaporate in close to a Planck time ($\sim 10^{-43}$s),
violating baryon number and any other global quantum numbers that might otherwise get in
the way of converting matter to radiation.  
For objects not bound to our galactic cluster, 
it is better to beam their energy back as radiation rather than as rest mass 
because the kinetic energy of a non-relativistic particle redshifts faster than the energy of radiation, 
so that one would have to continually supply additional energy to keep such particles on a trajectory 
that would intersect that of the central storage location.  
In addition, one may beam back energy from greater distances in this way.
If our expanding shell of micro-converter factories moves out at approximately the speed of light, 
then matter at a comoving distance $r$ will be encountered at a conformal time $\eta_e(r)=r/c$, 
and the corresponding radiation will be received by us at $r=0$ at $\eta_a(r)=2 r/c$.  
The ratio of the energy of the absorbed radiation to that emitted will be given by:
\be
\frac{E_a(r)}{E_e(r)} = \frac{1 - 2 a_0 H r/c}{1 - a_0 H r/c}
\ee
                                             
\begin{figure}[t]
\label{fig1}
\includegraphics[width=0.85\linewidth]{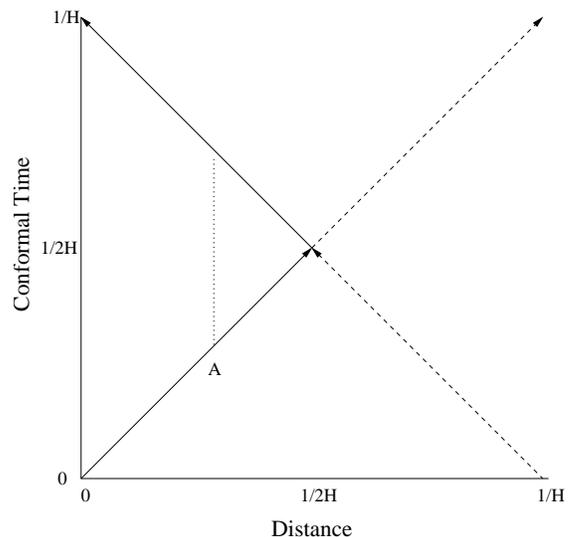}\\
\caption{
Light-cone coordinate representation of the (conformal) future of the universe in de Sitter space, 
showing the future event horizon, 
and the maximum coordinate distance to which an observer who leaves the origin at $t=0$ can travel 
and still remain in causal contact with the origin.  
Also shown, is the maximal conformal time at which a computer at rest at point A 
can process information and still communicate results back to the origin.
}
\end{figure}

In terms of a distance parameter $R=a_0 r$, 
the maximum amount of energy that will be received at
the origin by harvesting matter out to $R=R_\star$ is given by 
\bea
\label{Emax}
E_{\rm max}(R_\star) 
&= \int_0^{R_\star} 4\pi R^2 dR \rho_M(R) \frac{E_a(R)}{E_e(R)} \\
&= \frac{4\pi}{3}R_\star^3 \left(\Omega_m\rho_c\right)\left[1 - \left(H R_\star/c\right)\right]^3 ,\nonumber 
\eea
where $\rho_m$ is the energy density of matter in the universe, 
$\rho_c$ is the total energy density and $\Omega_m=\rho_m /\rho_c$.  
Note that $E_{\rm max}(R_\star)$ has a maximum at $HR_\star/c=1/2$.  
This is easily understood physically, as shown in Figure 1.  
With the de Sitter horizon located at $HR/c=1$,
the future light-cone of all objects that are sent away from the origin at $R=0$ today 
will intersect with $R=0$ in the future only if $HR/c \leq 1/2$.   
Thus, integrating over distances larger than this is unphysical.
Substituting this maximum value of $R_\star$ in equation (\ref{Emax}) yields.
\be
\label{Emaxb}
E_{\rm max} = \frac{\Omega_mc^5}{128 G H} .
\ee
The accessible energy thus works out to be $1/64^{\rm th}$ 
of the total energy located within the de Sitter horizon at the present time. 
Only $1/8{\rm th}$ of the commoving horizon volume is accessible 
for transporting energy back to the origin at the speed of light, 
and the mean redshift factor for the transported energy density 
is $1/8^{\rm th}$ that in a pure de Sitter universe.  
This upper limit is not only optimistic, it is also only exact in the limit of pure de Sitter expansion. 
If we relax this assumption, 
we can perform an analytic approximation as a power series in $\Omega_m/\Omega_\Lambda$,
the ratio of matter energy density to vacuum energy density. To first order this yields
\be
E_{\rm max} = \frac{\Omega_mc^5}{128 G H} \frac{1}{\Omega_\Lambda^3}
\left( 1 - \frac{3\Omega_m}{4\Omega_\Lambda}\right) .
\ee
For a matter density today of $30$\% of closure density, 
this would increase the net energy harvested by a approximately a factor of $2$.
The numerical magnitude of the total energy is also interesting.  
Using the current measured value of the Hubble constant, $H \approx 70{\rm km}/{\rm s}/{\rm Mpc}$,
one finds $E_{\rm max} \approx 3.5\times10^{67}$J, 
comparable to the total rest-mass energy of baryonic matter within today's horizon.  

This total accessible energy puts a limit on 
the maximum amount of information that can be registered and processed at the origin 
in the entire future history of the Universe.  
We can estimate this in either of two ways, both of which lead to the same value, 
and both of which also depend upon the acceleration of the Universe.  
In a de Sitter universe, the existence of an event horizon at $c/H$ 
implies that observers will be bathed in Hawking radiation with temperature $\hbar H/2\pi k_{\rm B}$.  
One cannot cool systems below this temperature without a constant expenditure of energy.   
Thus we expect that our computer will run at a temperature in excess of this temperature. 
However, to process information down a channel with noise temperature $T$ 
requires a minimum energy loss of  $k_{\rm B}T \ln 2$.\cite{Levitin}  
Dividing the total energy  (\ref{Emaxb}) by this value 
yields a limit on the number of bits that can be processed at the origin 
for the future of the Universe:
\be
\label{InfoProcessed}
{\rm Information \ Processed } \leq \frac{\pi \Omega_m c^5}{64 \hbar G H^2 \ln 2} = 1.35\times10^{120} . 
\ee

Precisely the same value is obtained by simply considering 
the maximal entropy associated with the energy harvested at the origin (obtained by partitioning the available energy in modes associated with the Hawking-Bekenstein temperature).  
The information that can be registered by a system is $I= S \ln 2$.  
If one assigns a temperature of  $\hbar H/2\pi k_{\rm B}$ to the above system, 
then the value one obtains for the information that can be registered equals that given in 
(\ref{InfoProcessed}).

It is interesting that the numerical value in (\ref{InfoProcessed}) 
for the future information processing capacity of an observer in an accelerating Universe 
is comparable to the value claimed for the computational capacity of our entire observable Universe 
over its past history\cite{Lloyd2002}.
One might wonder whether one could avoid the limit (5) by considering a distributed computing network, 
in which the machines deposited at each shell harvest the matter in their vicinity 
and compute directly with that material, 
ultimately sending the results of their computations back to the origin for final processing.   
This has the advantage of avoiding the redshift factor $E_a/E_e$R in (\ref{Emax}); 
but, as noted previously, each of these processors 
will only be able to perform operations for a finite time 
before it loses causal contact with an observer at the origin.  
At a distance $r=cx/H$, one finds an allowed processing time of 
\be
t_{\rm processing}(x) =  H^{-1}\log\left[\frac{1-x}  {x}\right]
\ee
Given the universal limit\cite{Margolis} of $2E/\pi\hbar$ 
on the number of operations per second that can be performed by a system of energy $E$, 
one can integrate over the total number of operations that can be performed by all processors
to determine the maximum number of distributed operations:
\bea
\int_0^{1/2} \frac{2 c^3}{\pi \hbar} \left[\frac{4\pi x^2 dx}{H^3} \Omega_m \rho_c (1-x)^3\right]
     H^{-1}\log\left[\frac{1-x}  {x}\right]  \cr
= \frac{\Omega_m \rho_c c^3}{\hbar H^4}\left[\frac{2}{15}\log 2 - \frac{17}{720}\right].
\quad\quad\quad\quad\quad\quad\quad\quad\quad\quad 
\eea
If one assumes that each operation processes at most on order of $1$ bit,
then the total bits processed in this way is approximately $1/6^{\rm th}$ 
of the total that can be processed directly at the origin, 
where one has an arbitrarily long time to process the stored energy.
It is remarkable that 
the effective future computational capacity for any computer in our Universe is finite, 
although, given the existence of a global event horizon, it is not surprising.  

Note that if the equation of state parameter $w$ for dark energy is less than -1 similar, implying that the rate of acceleration of the Universe increases with time, then similar although much more stringent bounds on the future computational capacity of the universe can be derived.  In this latter case, distributed computing is more efficient than local computing (by a factor as large as $10^{10}$ for $w -1.2$, for example), because the Hawking-Bekenstein temperature increases with time, and thus one gains by performing computations earlier in time.

The specific limit that we have derived here is of interest for several reasons.  
First, it demonstrates that the effective information available 
to any observer within the event horizon of an expanding universe 
is less than the Hawking-Bekenstein entropy.  
Second, the numerical value we have derived 
puts several fundamental limits on any future technological civilization.   
If consciousness involves information processing, 
then when one is ultimately able to determine the minimum complexity of a conscious being 
in terms of the information-processing rate in bits/sec, 
then an upper limit on the future of consciousness within an accelerating universe can be derived.  
On a more concrete level, perhaps, 
our limit gives a physical constraint 
on the length of time over which Moore's Law can continue to operate.  
In 1965 Gordon Moore speculated that the number of transistors on a chip, 
and with that the computing power of computers, would double every year\cite{Moore}. 
Subsequently this estimate was revised to between 18 months and 2 years, 
and for the past 40 years this prediction has held true, 
with computer processing speeds actually exceeding the 18 month prediction.   
Our estimate for the total information processing capability 
of any system in our Universe \ref{InfoProcessed}
implies an ultimate limit on the processing capability of any system in the future, 
independent of its physical manifestation and
implies that Moore's Law cannot continue unabated for more than 600 years 
for any technological civilization.

We acknowledge the hospitality of the Kavli Institute for Theoretical Physics 
during some of the period of this investigation.  LMK acknowledges M. Willbanks for 
calculations in his senior thesis confirming results regarding the $w <-1$ case.
Correspondence and requests for materials should be addressed to krauss@case.edu or glenn.starkman@case.edu

\end{document}